\documentclass[twocolumn,showpacs,preprintnumbers,amsmath,amssymb]{revtex4}

\usepackage{graphicx}
\usepackage{dcolumn}
\usepackage{bm}

\begin{document}

\preprint{B\"ottcher}

\title{On the proper reconstruction of complex dynamical systems spoilt by strong measurement noise}
\author{Frank B\"ottcher}
 \email{frank.boettcher@uni-oldenburg.de}
\author{Joachim Peinke}
 \email{peinke@uni-oldenburg.de}
\affiliation{Institute of Physics, University of Oldenburg,  D-26111
Oldenburg, Germany }

\author{David Kleinhans}
\author{Rudolf Friedrich}
\affiliation{Institute of Theoretical Physics, University of
M\"unster,  D-48149 M\"unster, Germany}

\author{Pedro G. Lind}
\author{Maria Haase}
\affiliation{University of Stuttgart, Pfaffenwaldring 27, D-70569
Stuttgart, Germany}

\date{\today}

\begin{abstract}
This article reports on a new approach to properly  analyze time
series of dynamical systems which are spoilt by the simultaneous
presence of dynamical noise and measurement noise. It is shown that
even strong external measurement noise as well as dynamical noise
which is an intrinsic part of the dynamical process can be
quantified correctly, solely on the basis of measured times series
and proper data analysis.  Finally  real world data sets are
presented pointing out the relevance of the new approach.
\end{abstract}

\pacs{05.10.Gg, 05.40.Ca}

\maketitle

\noindent A major challenge in analyzing time series originating
from complex systems is to reveal the underlying process dynamics.
Typically the simultaneous involvements of non-linearities,
dynamical noise and measurement noise cause problems for many
experimental situations and account for the complexity of this task.
The handling of these complications is the central concern of this
paper.

To extract an underlying signal disturbed by noise, linear and
non-linear predictor models or noise reduction schemes are widely
used (for discussion see \cite{kantz97} and references therein).
Here we choose an alternative approach based on the broad class of
Langevin processes which describes a variety of complex dynamical
systems.

Let us consider a one-dimensional Langevin process (the extension to
more dimensions is straightforward) that is given by:
\begin{eqnarray}
    \dot{x} = D^{(1)}(x) + \sqrt{D^{(2)}(x)}\;\Gamma(t) \;\;\; .
    \label{langevin}
\end{eqnarray}
The term $\Gamma(t)$ represents Gaussian white noise with
$\left<\Gamma(t)\right>= 0$ and
$\left<\Gamma(t')\Gamma(t)\right>=\delta(t-t')$. The terms
$D^{(n)}(x)$ are called drift coefficient ($n=1$) and diffusion
coefficient ($n=2$) and reflect the deterministic and the stochastic
part respectively. $\sqrt{D^{(2)}}$ fixes the amplitude of the
stochastic part and is referred to as {\it dynamical noise}. If
$D^{(2)}$ depends on $x$ it is called multiplicative noise;
otherwise it is called additive noise.

In recent years a parameter-free reconstruction of the coefficients
and thus of the corresponding Langevin process has been achieved
\cite{PRL97,siegert98,Ryskin97,friedrich00, gradisek00}. It has been
successfully demonstrated that traffic flow dynamics \cite{kriso02},
the chaotic dynamics of an electronic circuit \cite{siefert03,
siefert04} or the human heart beat rhythm \cite{kuusela04} can be
reconstructed without need of  any {\it a priori} models but just
from measured time series and the estimated drift and diffusion
coefficients. This estimation  is based on the evaluation of the
first ($n=1$) and the second ($n=2$) conditional moments:
\begin{eqnarray}
    M^{(n)}(x,\tau) &=& \left< \left[ x(t+\tau)-x(t) \right]^{n} \right>|_{x(t)=x}
    \label{moments}
\end{eqnarray}
from which the coefficients are derived according to:
\begin{eqnarray}
    D^{(n)}(x) &=& \lim\limits_{\tau \rightarrow 0}\;\; \frac{1}{\tau}M^{(n)}(x,\tau)   \;.
    \label{coefficients}
\end{eqnarray}
For ideal time series with a sufficient temporal resolution the
coefficients $D^{(n)}(x)$ can unambiguously be obtained from Eq.
(\ref{coefficients}). For real data sets however the sampling
frequency might be too low to resolve the dynamics properly as it
was pointed out in \cite{ragwitz01, sura02}. For small but finite
$\tau$ the conditional moments are better approximated by an {\it
Ito-Taylor series expansion} (e.g.
\cite{sura02,friedrich02,kloeden99}):
\begin{eqnarray}
     M^{(1)}(x,\tau) &\approx& \tau  D^{(1)}(x)+\mathcal{O}(\tau^{2})  \nonumber \\
    M^{(2)}(x,\tau) &\approx& \tau  D^{(2)}(x)
    +\mathcal{O}(\tau^{2}) \;.
    \label{correc0}
\end{eqnarray}
Depending on the process it might be necessary to consider further
higher order terms for estimating the coefficients. To finally
decide whether an obtained set of coefficients represents the real
dynamics at least a consistency check between the statistical
properties (moments, probability densities, etc.) of the
reconstructed and of the original times series has to be performed
(cf. \cite{renner01}).

Another important effect that complicates a proper estimation of
$D^{(n)}$ is the presence of {\it measurement noise}
$\sigma\zeta(t)$ (with $\left<\zeta(t)\right> = 0$ and
$\left<\zeta(t)\zeta(t')\right>=\delta(t-t')$) which is superimposed
on the data. Measurement noise corresponds to a rather unavoidable
experimental situation (e.g. \cite{kostelich93, kantz97, heald00})
and means that $y(t)=x(t)+\sigma\zeta(t)$ is examined rather than
$x(t)$. For instance, take the measurement of a turbulent velocity
time series. The resolution is chosen in such a way that the largest
fluctuations (on the largest time scales) are resolved. A certain
amount of measurement noise might be negligible for these large
scale fluctuations but can well be significant for the fluctuations
on the smallest scales where the fluctuations are much smaller! More
generally, the term 'measurement noise' refers to any superimposed
uncorrelated noise that is present in some complex system; it might
even be generated by the complex system itself.

To reconstruct the unknown dynamics $x(t)$ from the accessible
$y(t)$ it is thus essential to quantify $\sigma\zeta(t)$ and its
influence on the reconstruction of coefficients according to Eqs.
(\ref{coefficients}) and (\ref{correc0}), which will be the central
concern of this letter.

In \cite{siefert03, siefert04, renner01b} it has been shown that
measurement noise results in an offset term, $\gamma_{n}$, for the
conditional moments
\begin{eqnarray}
\label{offmeas}
    M^{(n)}(y,\tau) \rightarrow M^{(n)}(x,\tau) + \gamma_{n} \; ,\\
    \gamma_{1}=0\; ;        \; \gamma_{2}= 2\sigma^{2} \; .
    \label{offmalte}
\end{eqnarray}
Any non-zero offset causes a strong overestimation of the
coefficients, $D^{(n)}$, because it leads to a divergence of
$M^{(n)}(y,\tau)/\tau$ in Eq. (\ref{coefficients}). In
\cite{siefert03} it was therefore proposed to use the offset
$\gamma_{2}$ to quantify measurement noise and to take the slope of
the conditional moments (as a function of $\tau$) as an estimate of
the coefficients.

\begin{figure}[htbp]
\center
\begin{tabular}{cc}
\includegraphics[scale=0.29]{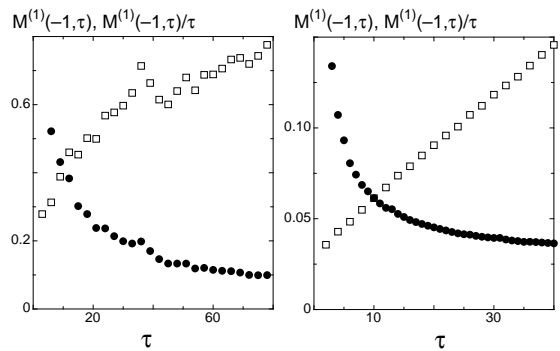}
\end{tabular}
\caption{\it Both plots show measured $M^{(1)}$ (squares) and
$M^{(1)}/\tau$ (circles) as a function of lag $\tau$. The left plot
refers to the North Atlantic Oscillation (NAO) index \cite{lind},
the right one to an increment time series of financial exchange
rates \cite{renner01b}. Units are arbitrary for clarity of
presentation. For both systems a significant offset of the first
conditional moment is observed causing a divergence of
$M^{(1)}/\tau$!} \label{realoff}
\end{figure}

In this paper we will show that Eqs. (\ref{offmeas}) and
(\ref{offmalte}) are restricted to the special case of low
measurement noise, that $M^{(1)}(y,\tau)$ must exhibit a
$\tau$-independent part (i.e. $\gamma_{1}\neq 0$) and that the slope
of the conditional moment, $M^{(n)}$, is no longer proportional to
the corresponding coefficient, $D^{(n)}$. In Fig. \ref{realoff} the
effect of measurement noise on $M^{(1)}$ is shown for two real world
examples, in which a strong offset at $\tau=0$ causes a divergence
of $M^{(1)}/\tau$. Finally, we propose an improved method to
quantify measurement noise even for very large noise levels. To this
end a general calculation of the conditional moments will be
performed. Thus we can explain the measurement noise dependence of
the second and the first conditional moment and propose an improved
reconstruction of the underlying process. For an Ornstein-Uhlenbeck
process the results are analytical, for non-Ornstein-Uhlenbeck
processes the corresponding analysis can be performed numerically.

Let us start with the calculation of the conditional moments from
the accessible data $y(t)$ spoilt by superimposed measurement noise.
Using the definition according to Eq. (\ref{moments}) the following
expressions can be derived:
\begin{eqnarray}
    \label{m1}
    M^{(1)}(y,\tau) &=& \left< y(t+\tau)-y(t)\right>|_{y(t)=y=x(t)+\sigma\zeta(t)} \nonumber \\
    &=& \tau \int D^{(1)}(x)f(x|y)dx + \int (x-y)f(x|y)dx \nonumber \\
    &=& m^{(1)}(y,\tau) + \gamma_{1}(y)\; , \\
    M^{(2)}(y,\tau)     &=&\tau \int \left[2(x-y)D^{(1)}(x) + D^{(2)}(x)\right]f(x|y)dx \nonumber \\
                    &+& \sigma^{2} + \int (x-y)^{2}f(x|y)dx\nonumber \\
                    &=&   m^{(2)}(y,\tau) + \gamma_{2}(y)\; .
    \label{m2}
\end{eqnarray}
The coefficients, $D^{(n)}$, are implicitly given by the conditional
moments that exhibit a $\tau$-dependent as well as a
$\tau$-independent part, denoted with $m^{(n)}(y,\tau)$ and
$\gamma_{n}(y)$, respectively. According to {\it Bayes' theorem} the
unknown probability density $f(x|y)$ is given by
$\frac{f(y|x)p(x)}{\int f(y|x)p(x)dx}$ where $f(y|x)$ denotes
nothing else than the distribution of measurement noise. Here we
consider Gaussian distributed measurement noise with variance
$\sigma^{2}$. The distribution of the process $x(t)$, given by Eq.
(\ref{langevin}), is denoted by $p(x)$. For stationary processes the
distribution is known to be
\begin{eqnarray}
    p(x)=\frac{\mathcal{N}}{D^{(2)}(x)} \cdot exp\left[\;\;2 \int\limits_{-\infty}^{x}
    \frac{D^{(1)}(x')}{D^{(2)}(x')}dx' \;\right]\;,
    \label{fp_solution}
\end{eqnarray}
where $\mathcal{N}$ denotes a proper normalization factor
cf. \cite{risken89}. To extract the coefficients from the four
equations of Eqs. (\ref{m1}) and (\ref{m2}) we assume for
convenience (but not necessarily) that the coefficients can be
modeled as polynomials. For instance, take the case of a
multiplicative process with $D^{(1)}=d_{11}x$ and
$D^{(2)}=d_{20}+d_{21}x+d_{22}x^{2}$. Then $5$ parameters ($\sigma,
d_{11}, d_{20}, d_{21}, d_{22}$) have to be derived by minimizing
the distance between the four measured functions,
$\hat{\gamma_{n}}(y), \hat{m}^{(n)}(y)$, and the solutions given by
Eqs. (\ref{m1}) and (\ref{m2}), i.e.
\begin{eqnarray}
    min\bigg\{ \sum\limits_{i}\left[ \hat{\gamma}_{1}(y_{i})-\gamma_{1}(y_{i}) \right]^{2}
    + \left[ \hat{\gamma}_{2}(y_{i})-\gamma_{2}(y_{i}) \right]^{2}+
    \nonumber \\
     \left[ \hat{m}^{(1)}(y_{i})-m^{(1)}(y_{i}) \right]^{2}
    + \left[ \hat{m}^{(2)}(y_{i})-m^{(2)}(y_{i}) \right]^{2}
    \bigg\}.
\end{eqnarray}\\
For an Ornstein-Uhlenbeck process and for pure noise, Eqs.
(\ref{m1}) and (\ref{m2}) can even be solved analytically. For the
latter case (i.e. $y(t)=\sigma\zeta(t)$) the moments are
given by:
\begin{eqnarray}
    M^{(1)}(y,\tau)&=& -y \nonumber \\
    M^{(2)}(y,\tau)&=& y^{2}+\sigma^{2} \; .
    \label{y}
\end{eqnarray}
This means that for pure noise the moments as a function of $\tau$
have vanishing slope but non-zero offset while for an ideal process
according to Eq. (\ref{langevin}) the situation is reversed. Data
from real processes will generally lead to $M^{(n)}$ values with
non-zero offsets and non-zero slopes.

Next we consider an  {\it Ornstein-Uhlenbeck process} (given by
$D^{(1)}=-\alpha x$ and $D^{(2)}=\beta$) to which measurement noise
is added. In this case $p(x)$ is a Gaussian distribution with zero
mean and variance $s^{2}=\beta/(2\alpha)$. The offsets
\begin{eqnarray}
    \gamma_{1}(y)&=&-\frac{\sigma^{2}}{\lambda^{2}}\cdot y  \nonumber \\
    \gamma_{2}(y)
    &=& \sigma^{2} +
    \frac{\sigma^{2}s^{2}}{\lambda^{2}}+\frac{\sigma^{4}}{\lambda^{4}}\cdot
    y^{2}
    \label{off}
\end{eqnarray}
and the $m^{(n)}$ values
\begin{eqnarray}
    m^{(1)}(y,\tau) &=& \tau \left[-\alpha y- \alpha \gamma_{1}(y)\right]  \nonumber \\
    m^{(2)}(y,\tau) &=& \tau \left[ \beta -2\alpha \left((\gamma_{2}(y)-\sigma^{2})+y\gamma_{1}(y)\right)\right]
    \label{M1}
\end{eqnarray}
can be derived exactly from Eqs. (\ref{m1}) and (\ref{m2}) (see
\cite{boettcher05} for details). Note that
$\lambda^{2}:=s^{2}+\sigma^{2}$ has been used and that $\gamma_2$
approaches $2\sigma^{2}$ in the small $\sigma$-limit in accordance
with Eq. (\ref{offmalte}).

From Eq. (\ref{M1}) it is seen that the slopes of the moments,
$m^{(n)}/\tau$, are affected by $\gamma_{n}$. Thus simply taking the
slope as an estimate of the coefficients
-- as suggested by Eq. (\ref{offmalte}) -- is not appropriate in
presence of larger measurement noise, even for rather simple cases
such as the Ornstein-Uhlenbeck process. Estimates according to Eq.
(\ref{coefficients}) will be increasingly in error as
$\gamma_{n}(y)$ dominates the conditional moments $M^{(n)}(y,\tau)$
for large $\sigma$.

\begin{figure}[htbp]
\center
\begin{tabular}{ll}
\includegraphics[scale=0.28]{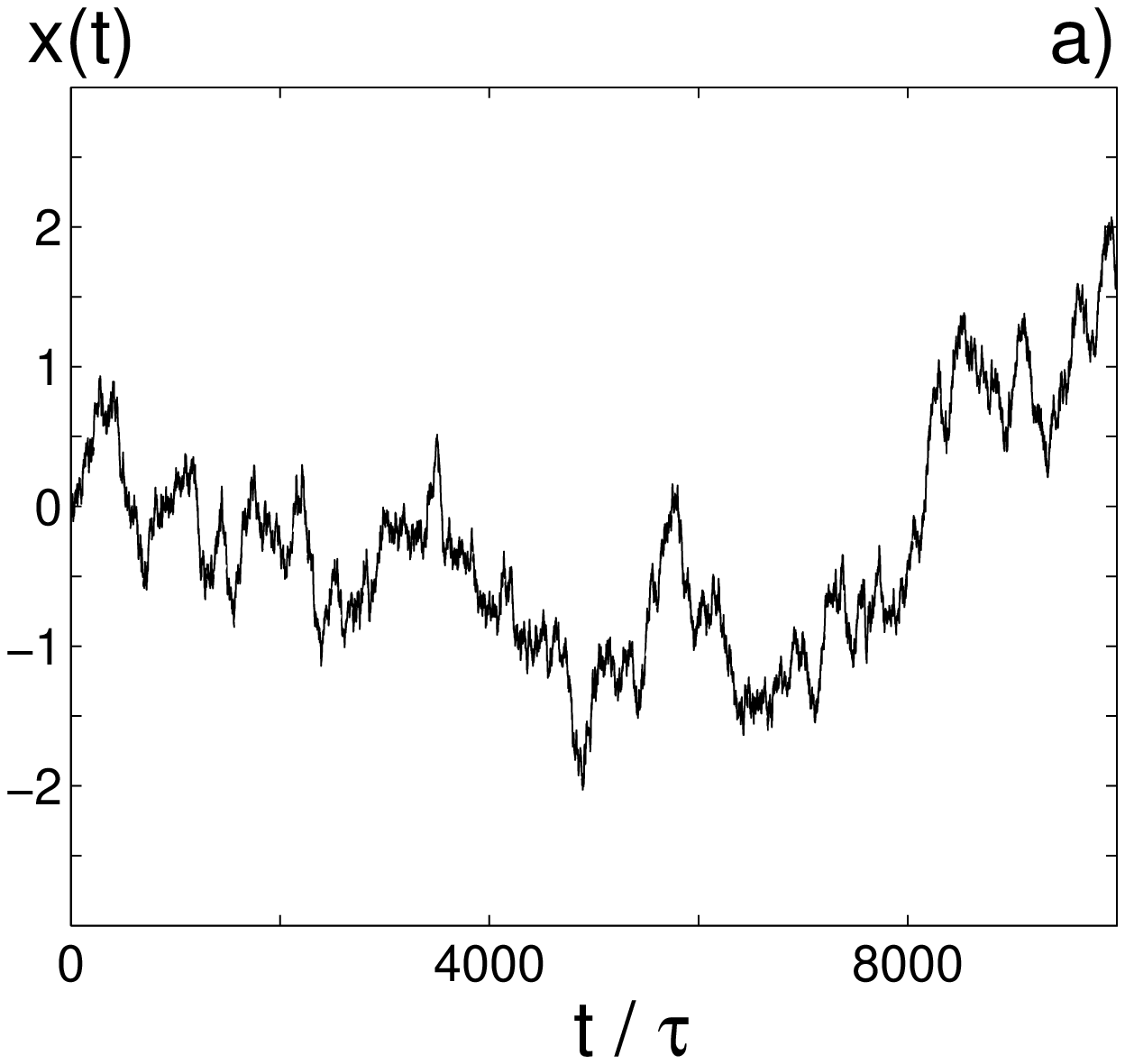} &
\includegraphics[scale=0.28]{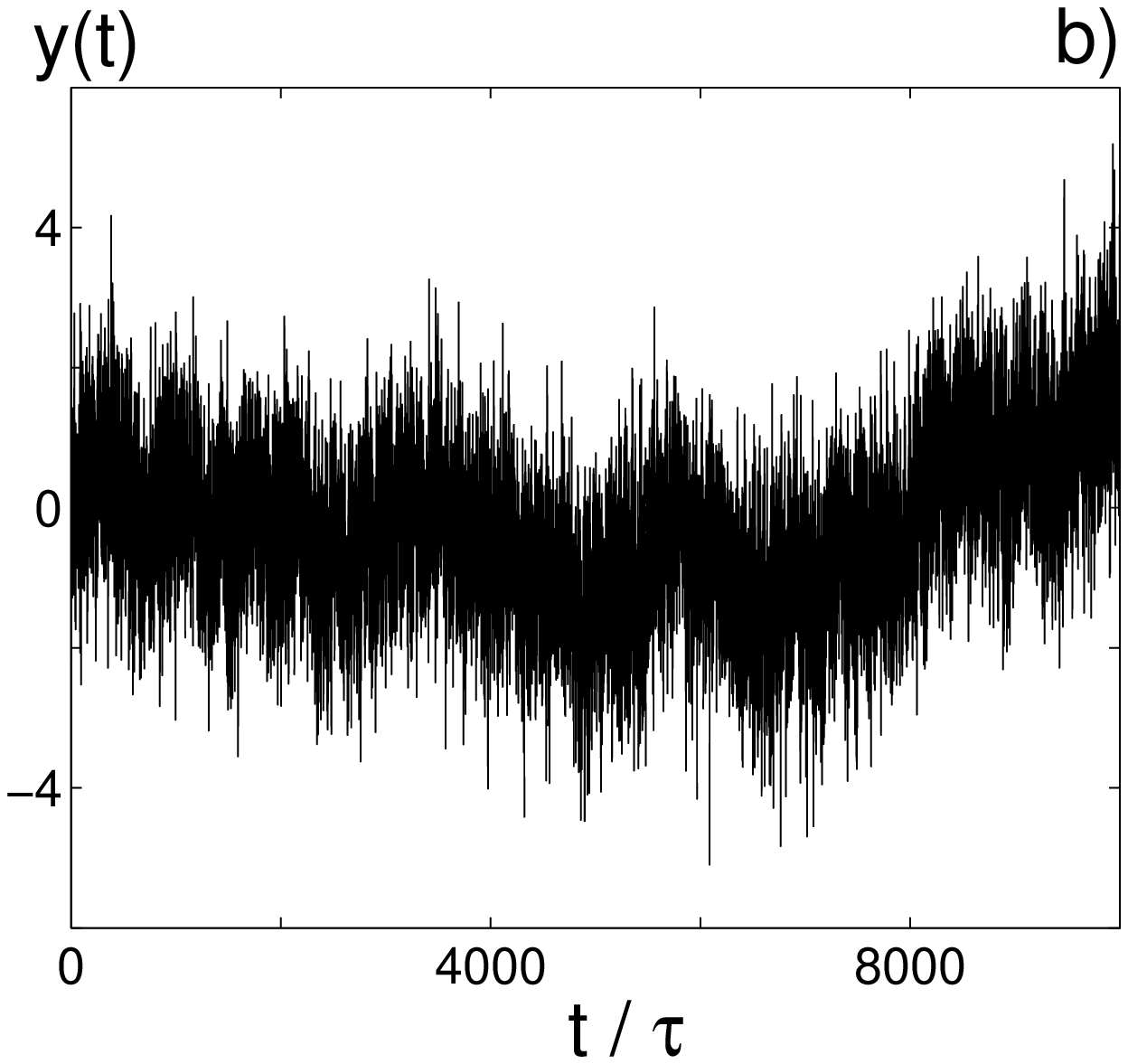}\\
\includegraphics[scale=0.29]{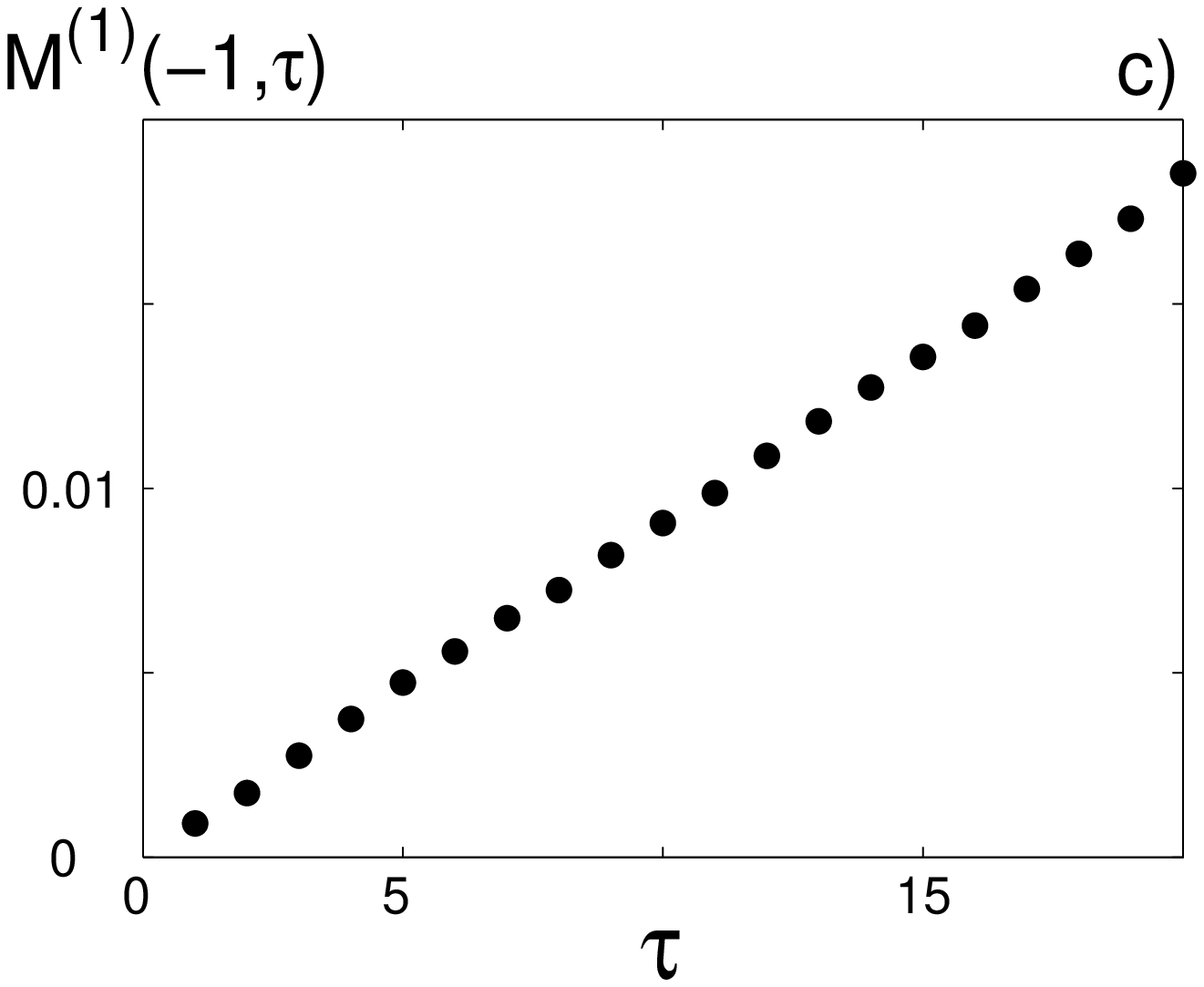} &
\includegraphics[scale=0.275]{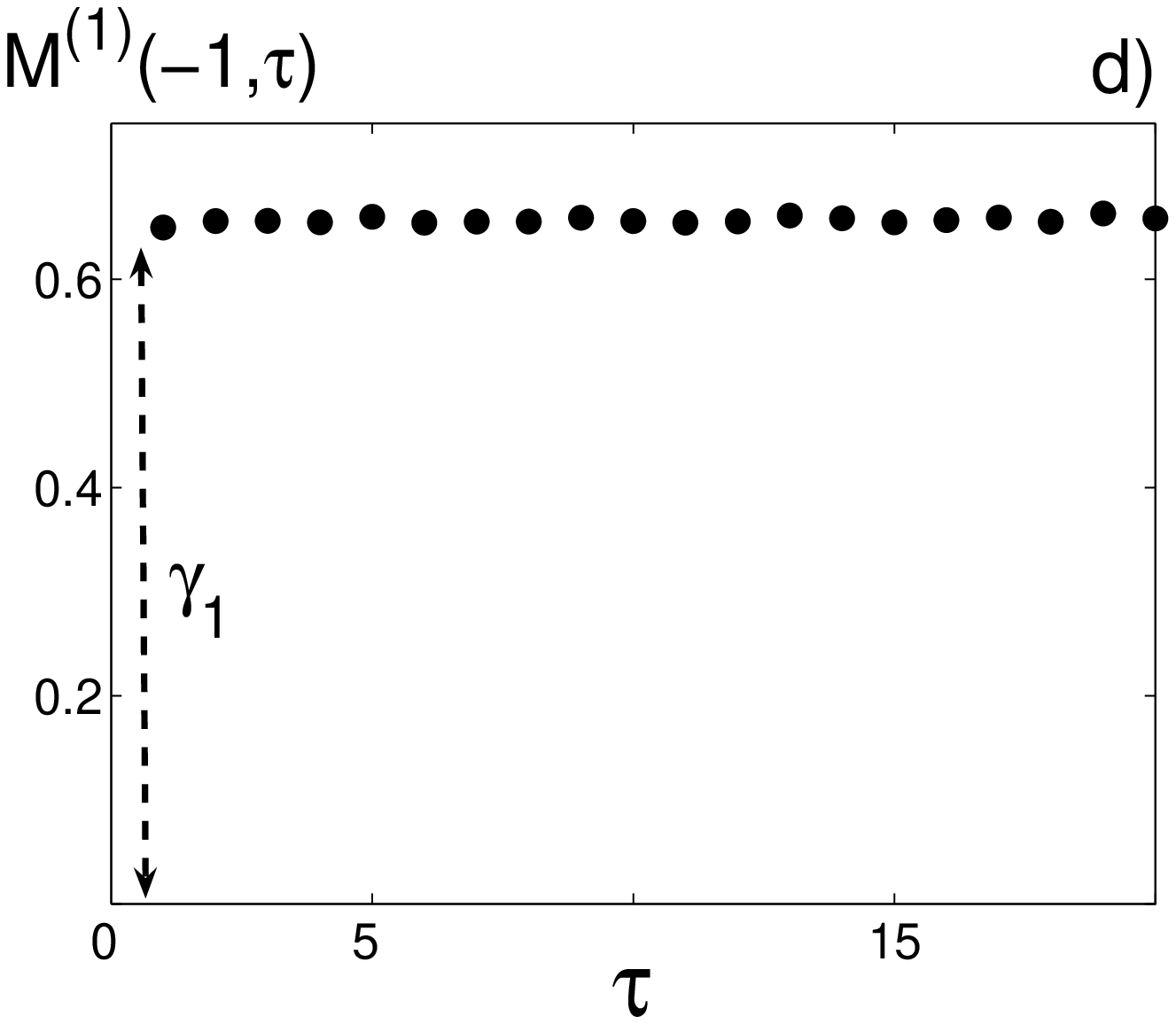}\\
\includegraphics[scale=0.3]{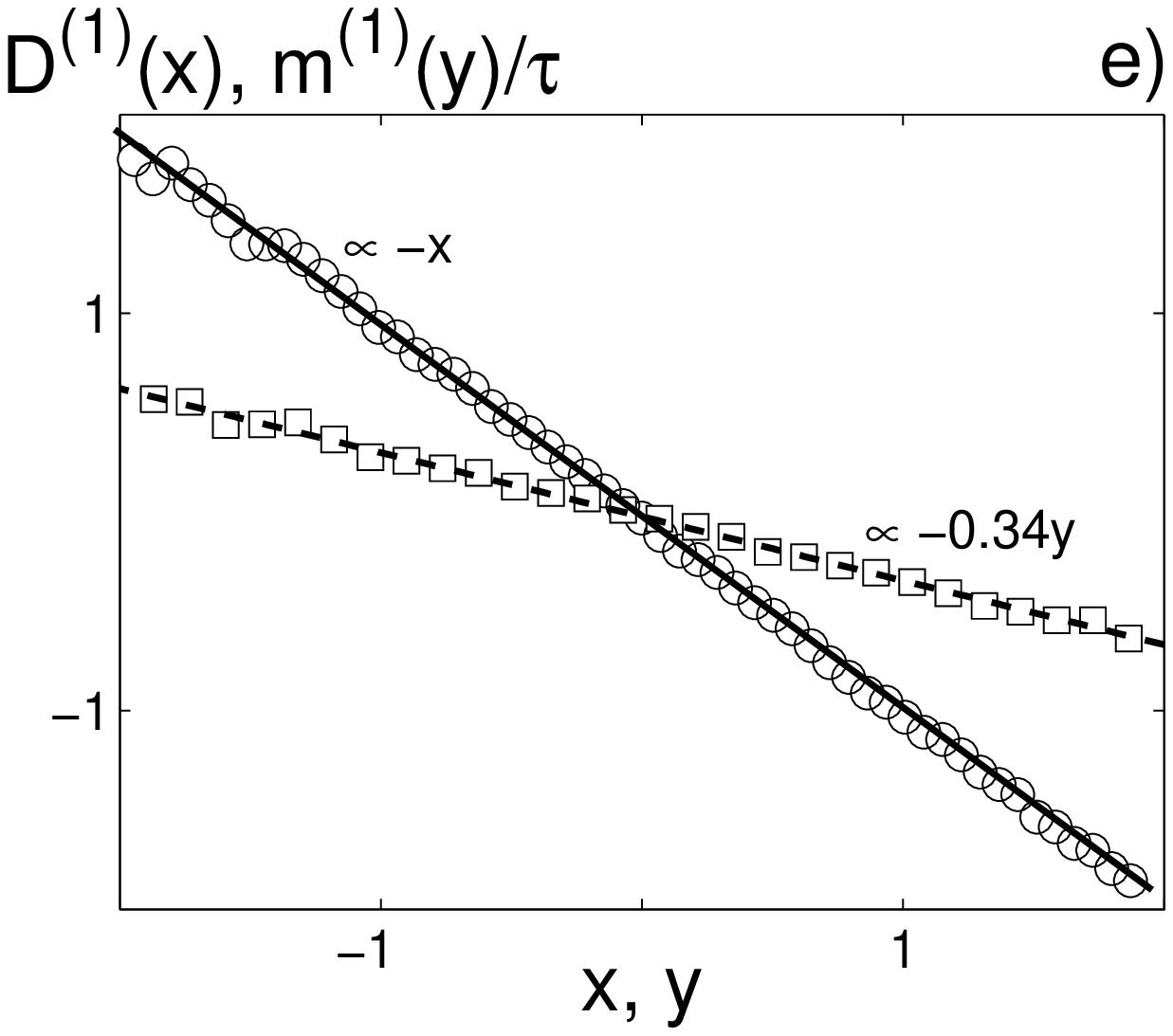} &
\includegraphics[scale=0.295]{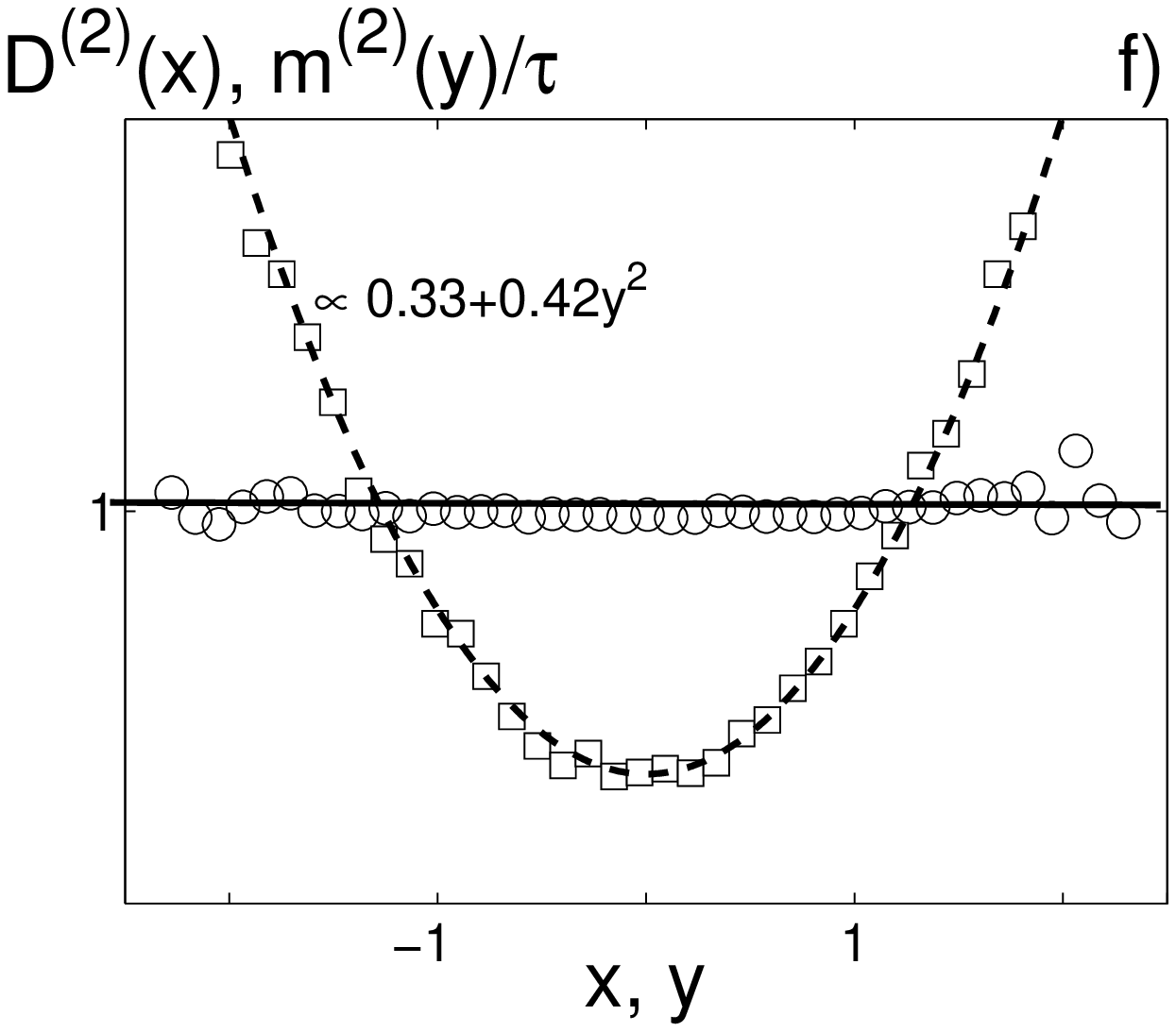}
\end{tabular}
\caption{\it a) and b) show an excerpt of an Ornstein-Uhlenbeck
process ($\tau=10^{-3}$, $\alpha=1$ and $\beta=1$) with $\sigma=0$
in a) and $\sigma=1$ in b). In c) and d) the derived $M^{(1)}$ are
shown, where c) refers to $x=-1$ and $\sigma=0$, and d) to $y=-1$
and $\sigma=1$. In e) and f) the symbols represent the measured
$D^{(n)}$ -- fitted by the solid lines ($\sigma=0$) and
$m^{(n)}/\tau$ -- fitted by the dashed lines ($\sigma=1$) according
to Eq.(\ref{M1}).} \label{plot}
\end{figure}

For illustration let us consider a numerical realization of an
Ornstein-Uhlenbeck process with $\alpha = \beta = 1$. Fig.
\ref{plot} a) shows the pure process ($\sigma=0$) and Fig.
\ref{plot} b) shows the process with strong superimposed measurement
noise ($\sigma = 1$), corresponding to a negative signal-to-noise
ratio of approximately $S/N=20\;log_{10} (s/\sigma)=-3dB$.
Without measurement noise the coefficients are directly obtained
either from the slopes of the conditional moments or by using Eq.
(\ref{coefficients}) as shown in previous works. The reconstructed
drift coefficient of Fig. \ref{plot} e) is found to be $\alpha=1\pm
0.01$ and analogously $D^{(2)}=\beta=1\pm 0.01$ is well
reconstructed (see Fig. \ref{plot} f)).

In presence of measurement noise the moments $M^{(n)}(y,\tau)$ are
still linear functions of $\tau$ but, in agreement with Eqs.
(\ref{m1}) and (\ref{m2}), exhibit an additional offset term as can
be seen in Fig. \ref{plot} d). From the measured $M^{(n)}(y,\tau)$
the terms $m^{(n)}(y,\tau)$ and $\gamma_{n}(y)$ are obtained as
follows:
\begin{eqnarray}
    m^{(1)}/\tau &=& -(0.34\pm 0.02)\cdot y  \nonumber \\
    m^{(2)}/\tau &=&  (0.33\pm0.02) + (0.42\pm0.01) \cdot y^2 \nonumber \\
    \gamma_1&=& (0.667 \pm 0.001)\cdot y \nonumber \\
    \gamma_2&=& (1.33 \pm 0.02)+ (0.445\pm0.002) \cdot y^2
    \label{all}
\end{eqnarray}
Using Eq. (12) we obtain the drift coefficient $D^{(1)}(x)=-\alpha
\cdot x$ with
 $\alpha = 1.01 \pm 0.02$ in good
agreement with the expected value of $\alpha=1$.

To reconstruct the diffusion term $D^{(2)}=\beta$ the knowledge of
$\gamma_{1}(y)$ and $\gamma_{2}(y)$ even at a single position $y$ is
sufficient when $\alpha$ is known. For instance, for $y=-1$ the
measured offsets are $\gamma_{1}=0.65$ and $\gamma_{2}=1.74$,
leading to $s=0.73$ and $\sigma=0.99$. With
$s=\sqrt{\beta/(2\alpha)}$ it follows that $\beta=1.01 \pm 0.04$. To
improve the accuracy of the parameters a least squares algorithm is
applied.

Based on the foregoing discussion of an Ornstein-Uhlenbeck process,
two important new results can already be given. Firstly, we see that
the estimation of the magnitude of measurement noise by the simple
approach according to Eqs. (5) and (6) is misleading. For instance
from the offset $\gamma_2(y=0)=2\sigma^2 \approx 1.34 \pm 0.02$ a
$\sigma^{2}$ value of $0.67$, which is about 67\% of the real value,
would be extracted. This underestimation has already been reported
in \cite{siefert03} and can now be understood quantitatively.
Secondly, if the small-$\tau$-estimate $m^{(2)}/\tau$ is taken as an
approximation of $D^{(2)}$, as it is commonly  done, then an
artificial quadratic diffusion term (see Fig. \ref{plot} f)) is
obtained, masquerading as multiplicative noise or a bad temporal
resolution \cite{sura02}.

Finally we consider a {\it general non-Ornstein-Uhlenbeck process}
where the coefficients, $D^{(n)}$ are implicitly given only by
$m^{(n)}$ and $\gamma_{n}$ according to Eqs. (\ref{m1} --
\ref{fp_solution}). As an example, let us consider a process with
multiplicative noise ($D^{(2)}=b+cx^{2}$) and linear drift
($D^{(1)}=-a x$) which is observed in various systems ranging from
finance to turbulence (cf. \cite{renner01, waechter04, tutkun04}).
Here we take $a=1$, $b=0.1$ and $c=0.5$ and iterate numerically. The
measurement noise amplitude is $\sigma\approx 0.25$ which
corresponds to a signal-to-noise ratio of $S/N=0$.
\begin{figure}[htbp]
\begin{center}
\begin{tabular}{cc}
\includegraphics[scale=0.25]{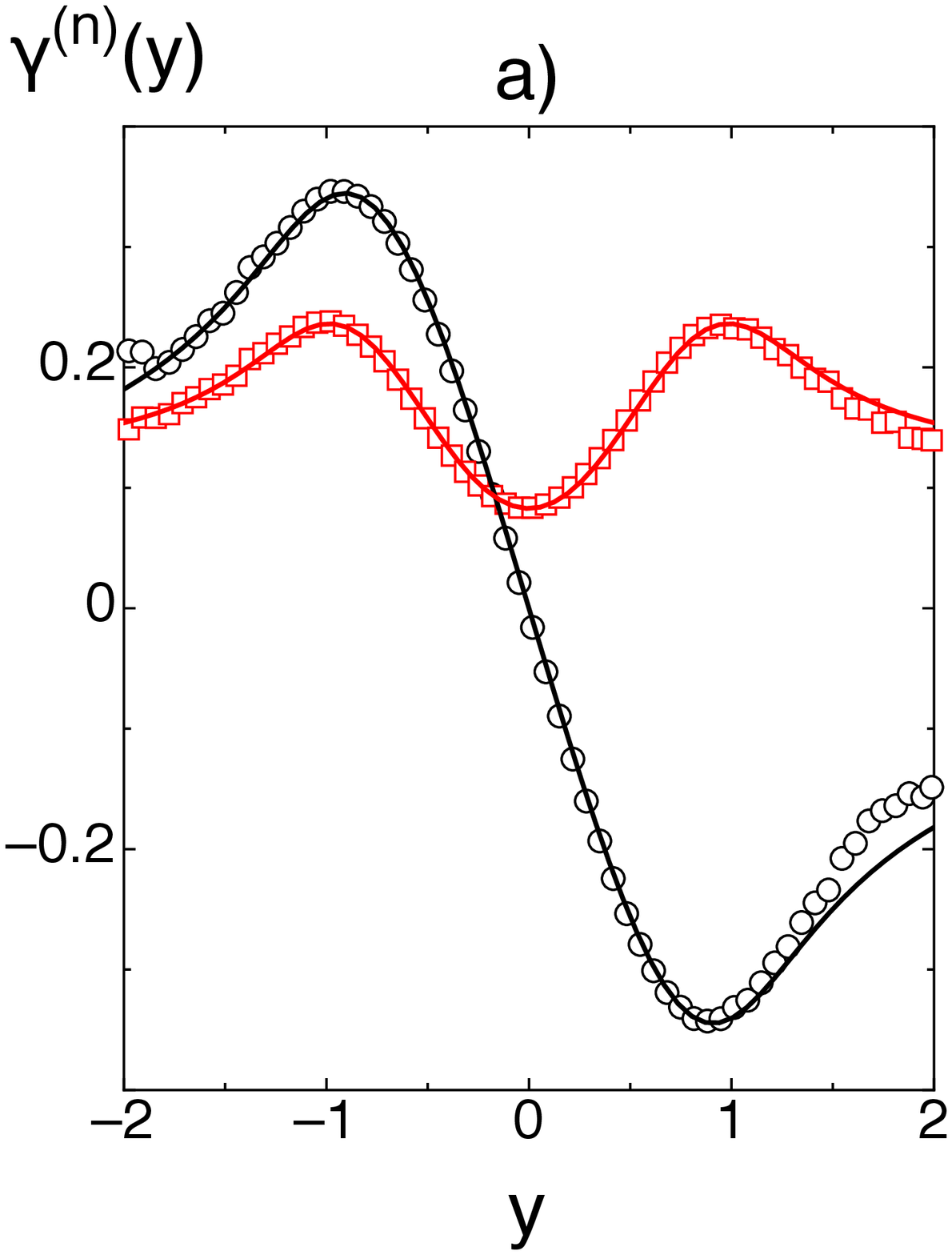}
& \includegraphics[scale=0.25]{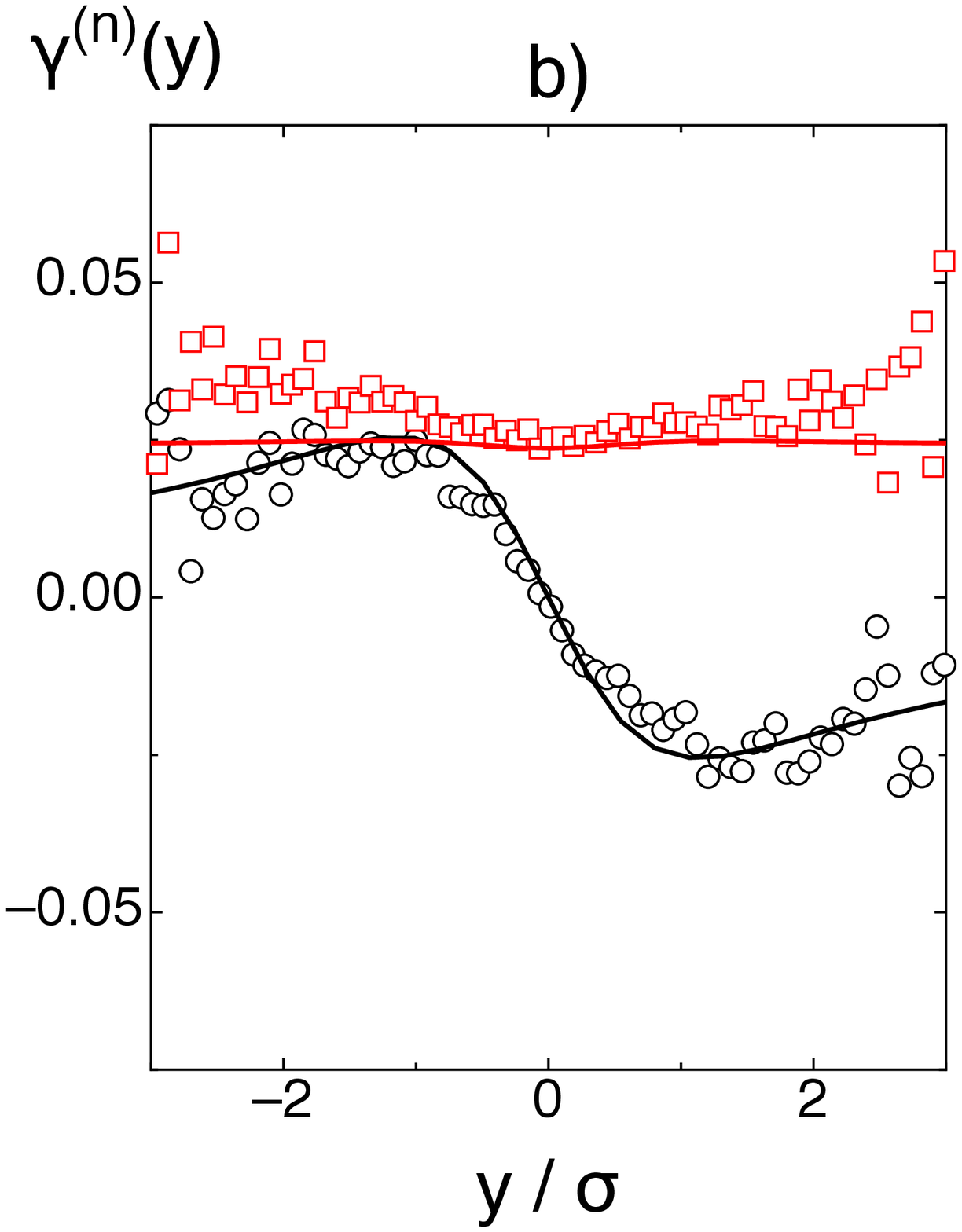}
\end{tabular}
\end{center}
\caption{\label{multi}  \it a) Symbols represent the measured
offsets of the iterated multiplicative process ($D^{(1)}=-x$,
$D^{(2)}=0.1+0.5x^{2}$ and $\sigma=0.25$). Solid lines represent the
expected offsets according to Eq. (\ref{m1}) and (\ref{m2}). b)
Measured offsets and numerical solutions for the financial data set.
In both plots circles refer to $\gamma_{1}$ and squares to
$\gamma_{2}$.}
\end{figure}\\
Fig. \ref{multi} a) shows the observed offsets, $\gamma_{n}$,
together with the (numerical) solutions according to Eqs. (\ref{m1})
and (\ref{m2}) which agree well. Obviously the offsets are
significantly different from those of an Ornstein-Uhlenbeck process
(see Eq. (\ref{off})). Fig. \ref{multi} b) shows the observed and
reconstructed offsets derived from the above presented financial
data. Especially for $\gamma_{1}$, the multiplicative character of
the underlying process becomes evident. This result shows that
multiplicative dynamical noise causes intermittent heavy-tailed
volatility statistics in financial data as it was proposed, and this
is not due to a spurious effect of measurement noise, as for the
case of Fig. \ref{plot}f).

From the examples of Fig. \ref{multi} we see again that the
understanding of the influence of measurement noise on the
conditional moments, i.e. on $\gamma_{n}$ and $m^{(n)}/\tau$, is the
key to achieving a proper reconstruction of the underlying dynamical
process (including the contribution of dynamical noise and
measurement noise) from pure data analysis. Naively applying the
definition according to Eq. (\ref{coefficients}) will no longer be
appropriate as soon as measurement noise is present.

To conclude, we have shown (for numerical as well as real world
data) that adding measurement noise to signals generated from a
Langevin process leads to a fundamental modification of the data
analysis via the conditional moments. A general equation describing
this modification has been presented and for the class of
Ornstein-Uhlenbeck processes analytical results are given. This
makes it possible to extract the strength of measurement noise,
$\sigma$, the standard deviation of the underlying process, $s$, as
well as the drift and diffusion coefficients, $D^{(1)}$ and
$D^{(2)}$, rather precisely even in presence of very strong
measurement noise. It is noteworthy that the evaluation of the
process' coefficients is solely based on analyzing the conditional
moments, which are directly obtained from the time series without
any need of pre-manipulating (e.g. filtering, modeling) the data.

\begin{acknowledgments}
We wish to acknowledge stimulating discussions with M. Siefert, St.
Barth, M. H\"olling and A. Nawroth.
\end{acknowledgments}


\begin{thebibliography}{}
    \bibitem{kantz97} H. Kantz and T. Schreiber, {\it Nonlinear Time Series Analysis} (Cambridge University Press, 1997).
    \bibitem{PRL97} R. Friedrich and J. Peinke,
    Phys. Rev. Lett. {\bf 78}, 863 (1997).
    \bibitem{siegert98} S. Siegert, R. Friedrich and J. Peinke, Phys. Lett. A {\bf 243}, 275 (1998).
    \bibitem{Ryskin97} G. Ryskin, Phys. Rev. E {\bf 56} 5123 (1997).
    \bibitem{friedrich00} R. Friedrich, et.al.
    Phys. Lett. A {\bf 271}, 217 (2000).
    \bibitem{gradisek00} J. Gradisek, S. Siegert, R. Friedrich and I. Grabec, Phys. Rev. E {\bf 62}, 3146 (2000).
    \bibitem{kriso02} S. Kriso, J. Peinke, R. Friedrich and P. Wagner, Phys. Lett. A {\bf 299}, 287 (2002).
    \bibitem{siefert03} M. Siefert, A. Kittel, R. Friedrich and J. Peinke, Europhys. Lett. {\bf 61}, 466 (2003).
    \bibitem{siefert04} M. Siefert and J. Peinke, Int. J. Bifurcation and Chaos {\bf 14}, 2005 (2004).
    \bibitem{kuusela04} T. Kuusela, Phys. Rev. E {\bf 69}, 031916 (2004).
    \bibitem{ragwitz01} M. Ragwitz and H. Kantz, Phys. Rev. Lett. {\bf 87}, 254501 (2001).
    \bibitem{sura02} P. Sura and J. Barsugli, Phys. Lett. A {\bf 305}, 304 (2002).
    \bibitem{friedrich02} R. Friedrich, C. Renner, M. Siefert and J. Peinke, Phys. Rev. Lett. {\bf 89}, 149401 (2002).
    \bibitem{kloeden99} P.E. Kloeden and E. Platen, {\it Numerical Solution of Stochastic Differential Equations} (Springer, 1999).
    \bibitem{renner01} C. Renner, J. Peinke and R. Friedrich, J. Fluid Mech. {\bf 433}, 383 (2001).
    \bibitem{heald00} J.P.M. Heald and J. Stark, Phys. Rev. Lett. {\bf 84}, 2366 (2000).
    \bibitem{kostelich93} E.J. Kostelich and T. Schreiber, Phys. Rev. E {\bf 48}, 1752 (1993).
    \bibitem{renner01b} C. Renner, J. Peinke and R. Friedrich, Physica A {\bf 298}, 499 (2001).
    \bibitem{boettcher05} F. B\"ottcher, Ph.D. thesis (Oldenburg, 2005).
    \bibitem{risken89} H. Risken, {\it The Fokker-Planck Equation} (Springer, 1989).
    \bibitem{waechter04} M. Waechter, F. Riess, T. Schimmel, U. Wendt and J. Peinke, Eur. Phys. J. B {\bf 41}, 259 (2004).
    \bibitem{tutkun04} M. Tutkun and L. Mydlarski, New J. Phys. {\bf 6} (2004) 49.
    \bibitem{lind} P.G. Lind, A. Mora, J.A.C. Gallas and M. Haase, IJBC, accepted, (2006).
\end{thebibliography}
\end{document}